# Trion Valley Coherence in Monolayer Semiconductors

Kai Hao[1], Lixiang Xu[1], Fengcheng Wu[1], Philipp Nagler[2], Kha Tran[1], Xin Ma[1], Christian Schüller[2], Tobias Korn[2], Allan H. MacDonald[1], Galan Moody[3*], and Xiaoqin Li[1*]

[1]Department of Physics and Center for Complex Quantum Systems, University of Texas at Austin, Austin, TX 78712, USA.

[2]Department of Physics, University of Regensburg, Regensburg, Germany 93040.

[3]National Institute of Standards & Technology, Boulder, CO 80305, USA.

*e-mail: elaineli@physics.utexas.edu; galan.moody@nist.gov

**The emerging field of valleytronics[1-3] aims to exploit the valley pseudospin of electrons residing near Bloch band extrema as an information carrier. Recent experiments demonstrating optical generation[1–4] and manipulation[5–7] of exciton valley coherence (the superposition of electron-hole pairs at opposite valleys) in monolayer transition metal dichalcogenides (TMDs) provide a critical step towards control of this quantum degree of freedom. The charged exciton (trion) in TMDs is an intriguing alternative to the neutral exciton for control of valley pseudospin because of its long spontaneous recombination lifetime, its robust valley polarization, and its coupling to residual electronic spin[8]. Trion valley coherence has however been unexplored due to experimental challenges in accessing it spectroscopically. In this work, we employ ultrafast two-dimensional coherent spectroscopy to resonantly generate and detect trion valley coherence in monolayer $MoSe_2$ demonstrating that it persists for a few-hundred femtoseconds. We conclude that the underlying mechanisms limiting trion valley coherence are fundamentally different from those applicable to exciton valley coherence. Based on these observations, we suggest possible strategies for extending valley coherence times in two-dimensional materials.**

The energy extrema in the band structure of monolayer TMDs occur at the *K* and *K′* points at the hexagonal Brillouin zone boundary. Broken spatial inversion symmetry and strong spin-orbit interactions introduce several valley-contrasting physical properties that enable manipulation of the valley pseudospin degree of freedom (DoF). Most notably, electrons and holes residing at the *K* and *K′* valleys possess opposite spin, orbital magnetic moment, and Berry curvature[9]. This intrinsic link between the electronic properties and the valley index has led to novel concepts in the burgeoning field of valleytronics. A seminal example in monolayer and bilayer TMDs as well as bilayer graphene is the valley Hall effect, which provides an electrical probe of the valley index[10,11].

Due to a sizable bandgap, the valley DoF in TMDs can be conveniently addressed and controlled optically. As with any binary quantum DoF, the valley pseudospin can be modeled as a two-level system and represented using the Bloch sphere shown in Fig. 1a. Using right- and left-circularly polarized light, photo-excitation of electron-hole pairs (excitons) into the $K$ and $K'$ valleys is possible, generating valley polarization. Excitons can also be prepared into a quantum mechanical superposition between the valleys using linearly polarized excitation, creating valley coherence[1–4]. An upper limit on the exciton valley coherence time is imposed by its recombination lifetime, which is ultrafast (a few hundred femtoseconds) in monolayer TMDs due to the intrinsically large exciton oscillator strength[12,13]. Furthermore, electron-hole exchange leads to faster exciton valley decoherence than that imposed by exciton recombination[2,14,15]. A natural question arises: can one generate, detect, and manipulate the valley quantum DoF via a different optical transition?

The charged exciton (trion) is a promising alternative to the exciton due to its longer recombination lifetime, robust valley polarization, and coupling to the additional charge. In molybdenum-based monolayers, the lowest-energy trion resonance is an inter-valley singlet state consisting of an electron-hole pair in one valley and an electron in the lowest conduction band in the other valley with opposite spin (Fig. 1b)[16–20]. Even after the electron-hole pair has recombined, the remaining residual electron may maintain definitive valley index and spin orientation, making the trion an attractive candidate to manipulate the valley DoF[21,22,24]. Although linearly polarized photon emission has been used to conveniently monitor exciton valley coherence, the same technique is not applicable to trions as we explain below. Following linearly polarized optical excitation, a coherent superposition of two trion valley configurations is created as shown in Fig. 1b. Upon electron-hole recombination, one configuration becomes a σ+ photon and spin-down electron while the other becomes a σ- photon and spin-up electron (Fig. 1c). In contrast to the exciton, their linear superposition leads to a spin-photon entangled state $|\sigma^+\rangle|\downarrow\rangle + |\sigma^-\rangle|\uparrow\rangle$, which cannot be detected as a linearly polarized emission due to the orthogonal spin states. For this reason, trion valley coherence has rarely been discussed in the existing literature, leaving a gap in our fundamental understanding of valley quantum dynamics in two-dimensional materials.

Here, we provide a direct measurement of trion valley coherence in monolayer $MoSe_2$ using polarization-resolved two-dimensional coherent spectroscopy (2DCS), which is a three-pulse four-wave mixing technique[23]. A pair of optical pulses with opposite helicity resonantly creates a coherent superposition of two trions in opposite valleys. Trion valley coherence is then read out by detecting the nonlinear four-wave mixing signal generated by a time-delayed third optical pulse. We have obtained a complete set of measurements evaluating the trion and exciton valley coherence, intra-valley interband optical coherence, and recombination dynamics by choosing a variety of different polarization schemes and scanning pulse delays. This comprehensive picture of the intra- and inter-valley quantum coherent dynamics associated with trions in monolayer $MoSe_2$ offers critical guidance for valley pseudospin manipulation.

An optical microscope image of our $MoSe_2$ sample is shown in the left panel of Fig. 2a. The sample is obtained by mechanical exfoliation onto a transparent sapphire substrate used for linear and nonlinear optical spectroscopy experiments in transmission. A micro-photoluminescence intensity spectrum is presented in the right panel of Fig. 2a, which is acquired using a linearly polarized pump at 1960 meV (633 nm) with the sample under vacuum at 20 K. Pronounced narrow-linewidth exciton and trion resonances observed at ~1650 meV and ~1620 meV, respectively, verify the high quality of the sample. For the trion we find similar photoluminescence intensity for orthogonal, linearly polarized detection angles (data not shown), consistent with the expected negligible degree of linear polarization explained previously.

Optical 2DCS is a particularly powerful tool for probing the full coherent valley dynamics of excitons and trions in TMDs[12,24,25]. The technique employs a series of 30-fs pulses generated from a mode-locked Ti:sapphire oscillator at an 80 MHz repetition rate and with sufficient bandwidth to excite both the exciton and trion transitions. The laser output is split into four phase-locked pulses using a platform of nested Michelson interferometers, enabling precision stepping of the pulse delays with ~1 fs resolution[26]. Three of the pulses are focused to a ~30 μm spot and interact nonlinearly with the monolayer to generate a four-wave mixing signal $E_S(t_1, t_2, t_3)$. The signal field is spectrally resolved in transmission through heterodyne detection with a phase-locked local oscillator derived from the fourth pulse (Fig. 2b). The pump fluence at the sample is kept below ~1 μJ cm$^{-2}$ (~3×10$^{11}$ excitons cm$^{-2}$) to remain in the $\chi^{(3)}$ regime and to minimize Auger-type recombination dynamics[27]. The valley coherence dynamics are obtained by recording the signal and scanning the delay $t_2$ between the second and third pulses while keeping the delay $t_1$ between the first two pulses held fixed (Fig. 2c). The signal field is Fourier transformed with respect to $t_2$ to generate a two-dimensional coherent spectrum $\mathbf{E}_S(t_1, \hbar\omega_2, \hbar\omega_3)$, which correlates the zero-quantum energies (*i.e.* valley coherence and lifetime) of the system during the delay $t_2$ with the one-quantum (*i.e.* emission) energies of the system during the emission time $t_3$.

Both non-radiative valley coherence and population recombination dynamics evolve during the delay $t_2$ and influence the nonlinear signal. Separating these contributions is possible through careful selection of the helicity of the excitation pulses and detected signal. We first present valley coherence measurements in which the first and third pulses are left-circularly polarized (σ+) and the second pulse and detected four-wave mixing signal are right-circularly polarized (σ-). As illustrated in Fig. 2c, the first pulse creates a coherent superposition between the crystal ground state and the trion in the *K* valley. After a delay $t_1$ (set to zero in the present experiments), the second pulse drives the system into a non-radiative coherent superposition of the trion configurations in the *K* and *K'* valleys. The trion valley coherence evolves during the delay $t_2$, after which the third pulse converts the valley coherence to an optical coherence in the *K'* valley, which is then radiated as the four-wave mixing signal and detected through heterodyne spectral interferometry. We note that this particular polarization scheme of the interacting fields is inaccessible in the commonly used pump-probe technique. Via this polarization sequence, we are able to isolate valley coherence among all possible third order quantum mechanical pathways.

The resulting zero-quantum spectrum for cross-circular polarization is shown in Fig. 3a. The spectrum features two resonances at emission energies of ~1619 meV and ~1649 meV corresponding to the trion valley coherence (*T*) and exciton valley coherence (*X*), respectively. Because the transitions in the *K* and *K'* valleys are degenerate[5,6], the valley pseudospin vector does not precess about the Bloch sphere during $t_2$ and instead its magnitude decays exponentially with a valley decoherence rate $\gamma_v$ (or valley coherence time $\tau_v = \hbar/\gamma_v$). As a result, the resonances in Fig. 3a appear centered at the origin of the zero-quantum axis ($\hbar\omega_2$) with a half-width at half-maximum (HWHM) equal to $\gamma_v$. By analyzing the lineshapes in Fig. 3b and Fig. 3c, we find that the valley coherence time for the trion and exciton is $\tau_v^T = 230$ fs ($\gamma_v^T = 2.9$ meV) and $\tau_v^X = 140$ fs ($\gamma_v^X = 4.8$ meV), respectively.

The similar order-of-magnitude for the exciton and trion valley coherence times is a surprising result because they have very different intrinsic recombination and valley properties. The large exciton oscillator strength and nonradiative decay processes result in ultrafast population recombination with a lifetime $T_1^X$ = 210 fs, placing a firm upper limit on the exciton valley coherence time. Furthermore, electron-hole exchange interaction accelerates exciton valley decoherence by breaking the two-fold valley degeneracy through the Maialle-Silva-Sham mechanism[14,15]. Likewise, both ultrafast population relaxation and the electron-hole exchange lead to rapid exciton valley depolarization from an initial exciton valley polarization of 50% within ~350 fs under resonant excitation conditions (see Figs. 3d-3f and the Supplemental Information). In the case of trions, both population relaxation and valley depolarization are expected to be slower than those of excitons. In the sample studied in this work with a low doping density, our measurements indeed yielded a trion recombination $T_1^T$ = 4.7 ps, a higher degree of initial valley polarization (~75%), and a subsequent valley depolarization time of ~3 ps (see Figs. 3d-3f and the Supplemental Information). Slower trion valley polarization dynamics is expected, as we now explain. First, unlike excitons with zero center-of-mass momentum, the inter-valley trions at the *K* and *K'* valleys carry opposite center-of-mass momentum. Second, the electron-hole exchange interaction causes valley depolarization via annihilation of an electron-hole pair in one valley and creation of another pair in the other valley. This process would correspond to the formation of an intra-valley trion in MoSe$_2$ which has a higher energy. The momentum conservation and energy up-conversion requirements both reduce the trion inter-valley scattering rate.

The fact that the trion valley coherence time is nearly an order-of-magnitude shorter than its recombination lifetime and valley depolarization dynamics (see Supplemental Information) suggests that the trion is more susceptible to pure dephasing from interactions with its local fluctuating environment. We present additional experiments in which the first delay t$_1$ between the pulses is scanned while the second delay t$_2$ is held fixed. The nonlinear signal is Fourier transformed with respect to t$_1$, which produces a one-quantum spectrum **E**$_S$(ℏω$_1$, t$_2$, ℏω$_3$) that correlates the excitation and emission energies of the system during the delays t$_1$ and t$_3$. The resulting one-quantum spectrum is shown in Fig. 4a. The diagonal dashed line indicates excitation and emission at the same energy; the two diagonal peaks at ~1619 meV and ~1649 meV correspond to the nonlinear response associated with the trion and exciton transitions, respectively. The peaks are slightly elongated along the diagonal, which indicates moderate inhomogeneous broadening of excitons and trions. The HWHM of the lineshapes along the cross-diagonal direction[28], shown in Fig. 4b, is approximately equal to the homogeneous linewidth γ, which is inversely proportional to the intra-valley coherence time $T_2$. We measure for the trion and exciton $T_2^T = 510$ fs ($\gamma^T = 1.3$ meV) and $T_2^X = 470$ fs ($\gamma^X = 1.4$ meV), respectively[12].

These results confirm our expectations: exciton decoherence is recombination lifetime limited ($T_2^X \approx 2T_1^X$), whereas trion decoherence is dominated by pure dephasing from its environment ($T_2^T \ll T_1^T$). This observation supports the view that trion features in the optical properties of most TMDs should be viewed as coming from exciton dressing by a Fermi sea, with internal quantum fluctuations, rather than a single electron[29]. Alternatively, pure dephasing might also arise from scattering of trions from charged surface adsorbates, as well as defects or impurities both in the monolayer TMD and the substrate through the long-range screened Coulomb interaction. In a quasi-2D GaAs quantum well, the screened Coulomb interaction is stronger for the trion compared to the charge-neutral exciton[30], which is consistent with the presence (absence) of pure dephasing for the trion (exciton) observed here. A summary of the intra- and inter-valley coherent and incoherent dynamics is provided in Table S1, which highlights the key differences between the exciton and trion. From density matrix calculations of the valley dynamics for a simple three-level $V$-system[31,32], one would expect that the valley decoherence is related to the population relaxation and pure dephasing ($T_2^*$) through $1/\tau_v = 1/T_1 + 2/T_2^*$. While this simple model doesn't hold for the exciton due to additional decoherence from the electron-hole exchange interaction, it is entirely compatible with the singlet trion for which inter-valley exchange is suppressed. This relation supports our assertion that pure dephasing of the trion resonance itself is the dominant valley decoherence mechanism for the trion.

In summary, we have implemented optical 2DCS to generate and detect trion valley coherence in monolayer semiconductors, which had been previously unexplored. While the trion and exciton valley decoherence occur on a similar time scale of a few hundred femtoseconds, their underlying mechanisms are very different. In the case of excitons, valley decoherence is caused by the intrinsically fast electron-hole exchange interaction and interband recombination, leaving little room for future improvement. Our key observation that trion valley decoherence is primarily from pure dephasing from interactions with the surrounding environment suggests that the valley coherence time can be enhanced, possibly simply by lowering the density of carriers. Placing TMD monolayers on high quality insulating substrates such as hexagonal boron nitride and encapsulating the TMD will likely shield trions from pure dephasing induced by defects and impurities. More importantly, the excitation of valley selective trions serves to optically initialize the spin/valley index of free electrons in the monolayer TMD. A tantalizing hypothesis is that after electron-hole recombination, the residual electron bound to the trion composite quasiparticle could maintain its valley polarization and coherence. This is consistent with the nanosecond long spin coherence time[21,22] observed in monolayer TMDs and should be tested in future experiments. Adiabatic excitation schemes may further be applied to realize valley qubit rotations similar to coherent control experiments performed on electron spins in quantum dots[33,34].

**Acknowledgements** The theoretical and experimental collaboration is made possible via SHINES, an Energy Frontier Research Center funded by the U.S. Department of Energy (DoE), Office of Science, Basic Energy Science (BES) under award # DE-SC0012670. K.H., F.W., L.X., X.L., and A.H.M. have all received support from SHINES. A.H.M. also acknowledges support from Welch Foundation F-1473. X. Li also acknowledges the support from NSF EFMA-1542747, a Humboldt fellowship and Welch Foundation F-1662. P. Nagler, C. Schüller and T. Korn gratefully acknowledge technical assistance by S. Bange and financial support by the German Research foundation (DFG) via GRK 1570 and KO3612/1-1.

**Author Contributions** G.M. and X.L. conceived the concept. K.H. led the experimental effort. All co-authors at the University of Texas ran the experiments, acquired the data, and analyzed the results. P.N., C.S. and T.K. provided the sample. F.W. and A.H.M. provided theoretical support. G.M. and X.L. wrote the manuscript. All authors discussed the results and commented on the manuscript at all stages.

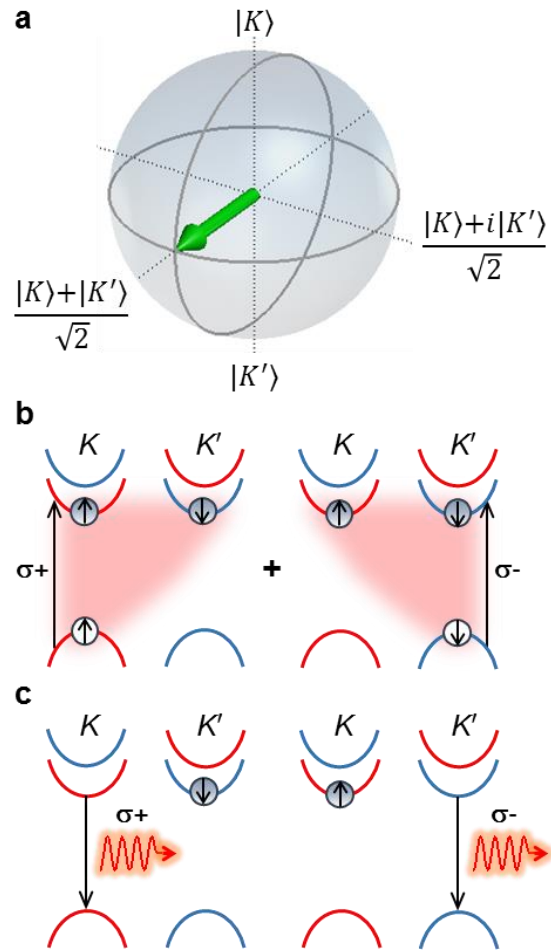

**Figure 1: Trion Valley Coherence in Monolayer MoSe$_2$.** **(a)** Valley polarization and coherence can be represented by a pseudospin vector on the Bloch sphere oriented along the north/south poles or in the equatorial plane, respectively. **(b)** Optical excitation using linearly polarized light can generate trion valley coherence between the lowest-energy negative trion states in MoSe$_2$. **(c)** Radiative recombination leads to a spin-photon entangled state due to the opposite residual electronic spins, which prevents detection of the trion valley coherence in linear polarized photoluminescence.

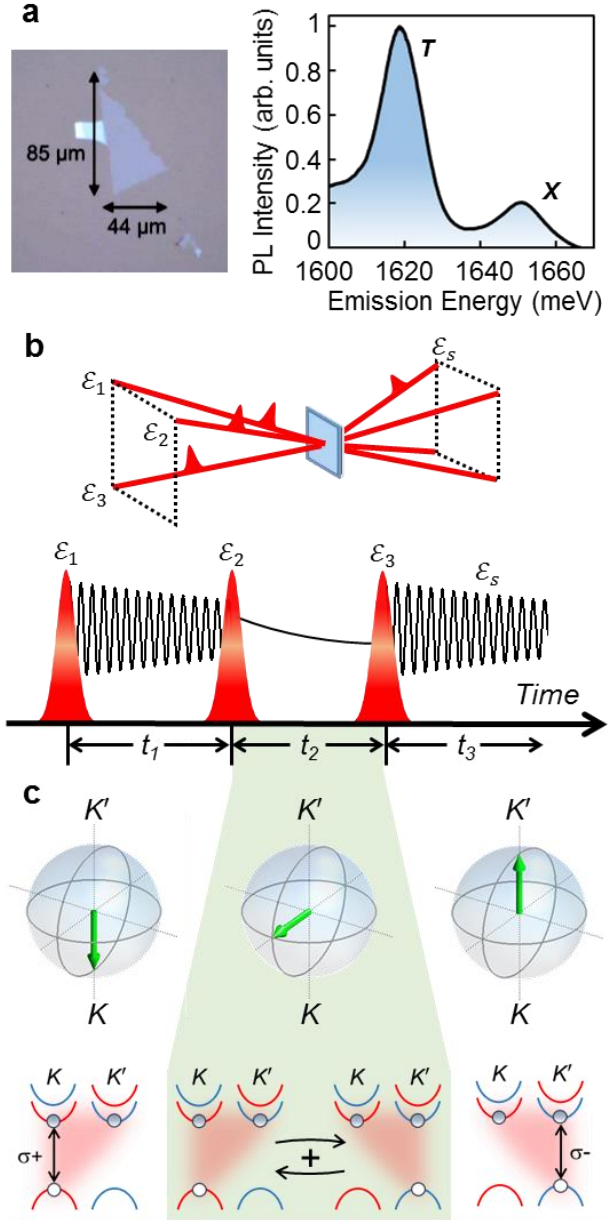

**Figure 2: Resonant Generation and Detection of Trion Valley Coherence.** **(a)** Optical microscope image of the MoSe$_2$ monolayer. The low temperature photoluminescence spectrum features two peaks corresponding to the trion and exciton transitions at ~1620 meV and ~1650 meV, respectively. **(b)** Optical 2DCS experiments are performed in the box geometry. Three pulses with variable delays and polarizations interact nonlinearly with the sample to generate a four-wave mixing signal that is detected in the phase-matched direction. **(c)** Cross-circularly polarized excitation and detection, in which the first and third (second and signal) pulses are left- (right-) circularly polarized. Generation and detection of valley coherence is illustrated by the Bloch vector.

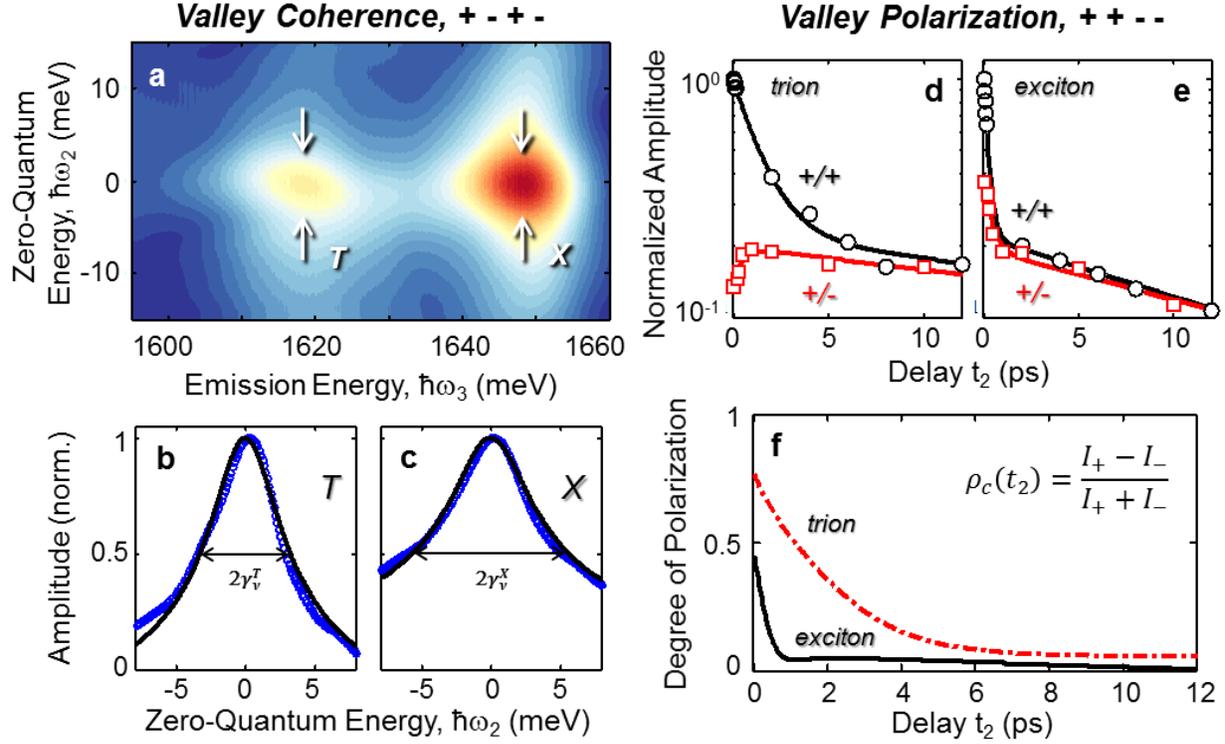

**Figure 3: Inter-Valley Coherence and Valley Polarization Dynamics.** (a) Optical 2DCS zero-quantum spectrum acquired using the cross-circular polarization scheme featuring two peaks at the emission energies of the exciton (*X*) and trion (*T*), respectively. (b)-(c) The half-width at half-maximum of Lorentzian fit functions (solid lines) to the lineshapes along the zero-quantum axis $\hbar\omega_2$ yields the valley decoherence rate $\gamma_v$ (valley coherence time $\tau_v = \hbar/\gamma_v$), equal to 2.9 meV (230 fs) and 4.8 meV (140 fs) for the trion and exciton, respectively. The valley polarization dynamics for the (d) trion and (e) exciton obtained from one-quantum spectra acquired for co- and cross-circularly polarized. (f) Bi-exponential fits to the amplitudes in (d) and (e) are used to determine the degree of circular polarization at various delays $t_2$.

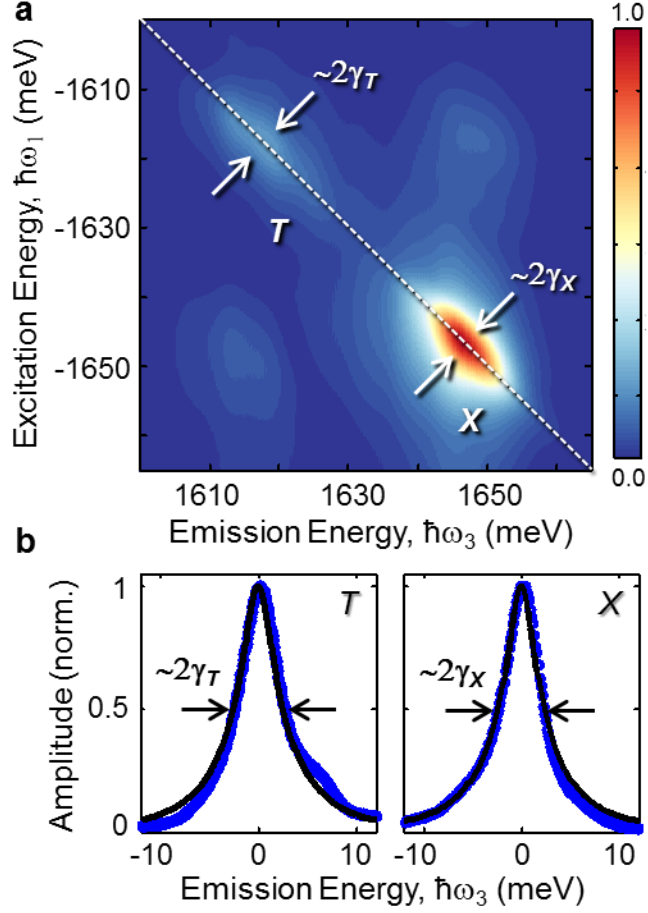

**Figure 4: Intra-Valley Coherence Dynamics. (a)** Optical 2DCS one-quantum spectrum acquired using a co-circular polarization scheme. The peaks on the diagonal dashed line correspond to excitation and emission at the exciton and trion resonances. **(b)** The resonance lineshapes along the cross-diagonal direction are fit with Lorentzian functions to obtain the homogeneous linewidth (coherence time) $\gamma^T$ = 1.3 meV ($T_2^T$ = 510 fs) and $\gamma^X$ = 1.4 meV ($T_2^X$ = 470 fs) for the trion and exciton, respectively.

# Supplementary Information for: Trion Valley Coherence in Monolayer Semiconductors

Kai Hao[1], Lixiang Xu[1], Fengcheng Wu[1], Philipp Nagler[2], Kha Tran[1], Xin Ma[1], Christian Schüller[2], Tobias Korn[2], Allan H. MacDonald[1], Galan Moody[3*], and Xiaoqin Li[1*]

[1] Department of Physics and Center for Complex Quantum Systems, University of Texas at Austin, Austin, TX 78712, USA.

[2] Department of Physics, University of Regensburg, Regensburg, Germany 93040.

[3] National Institute of Standards & Technology, Boulder, CO 80305, USA.

*e-mail: elaineli@physics.utexas.edu; galan.moody@nist.gov

## Valley Polarization Lifetime

The valley polarization dynamics are investigated by acquiring one-quantum spectra similar to the one shown in Fig. 4a, taken for various delays $t_2$. At each delay, a co-circularly polarized (+ + + +) and cross-circularly polarized (+ + - -) spectrum is recorded. In these types of experiments, because the first two pulses create exciton and trion populations, the valley dynamics of these populations can be extracted by comparing the relative amplitudes of the peaks in the co- and cross-circularly polarized spectra. The trion and exciton peak amplitudes versus delay $t_2$ are shown in Fig. 3d and 3e of the main text, respectively. The degree of circular polarization is calculated from these amplitudes as $\rho_c = (I_+ - I_-)/(I_+ + I_-)$.

We find that valley polarization for the trion and exciton approaches ~75% and ~50% at early times and exponentially decays within ~3 ps and ~350 fs, respectively. The degree of polarization in monolayer $MoSe_2$ observed here is nearly an order of magnitude larger than that obtained in experiments using non-resonant excitation[1]. The rapid decay of the exciton valley polarization and valley coherence can be attributed to the strong inter-valley exchange interaction, which acts as an effective in-plane magnetic field that breaks the *K* and *K'* degeneracy at finite momentum through the Maialle-Silva-Sham mechanism[2,3].

## Population Relaxation Dynamics

The exciton and trion recombination times are measured by repeating the zero-quantum spectrum experiments shown in Fig. 3 of the main text using co-circular polarization of the excitation pulses and detected signal. In this case, the first two pulses generate a population of excitons and trions in the K valley, which decay with rate $\Gamma$ (population decay time $T_1 = \hbar/\Gamma$) during the delay $t_2$. The third pulse converts the populations to an optical coherence in the same valley that radiates as the four-wave mixing signal. The resulting 2D spectrum is shown in Fig. S1a. Two peaks appear at the exciton and trion emission energies. Since the system is in a population state, the phase does not evolve during the delay $t_2$ leading to the peaks appearing at the origin of zero-quantum energy axis. The HWHM of the lineshape along the zero-quantum energy axis provides a measure of the population relaxation rates (Fig. S1b). The trion lineshape is fit with a single Lorentzian function with a width $\Gamma^T = 0.14$ meV ($T_1^T = 4.7$ ps), which is similar to previous measurements of the trion recombination lifetime[1,4]. The exciton is fit with a double Lorentzian function reflecting the few-hundred femtosecond and few-picosecond biexponential decay dynamics typically observed in monolayer TMDs[5,6]. The exciton population dynamics are dominated by the fast component with a decay rate $\Gamma^X = 3.2$ meV ($T_1^X = 210$ fs). In contrast, for the trion, the few-picosecond lifetime compared to the few-hundred femtosecond valley coherence time suggests that it is more susceptible to pure dephasing from interactions with its local fluctuating environment. A comparison of the various trion and exciton relaxation and coherence times is provided in Table S1.

**Table S1. Trion and Exciton Relaxation and Coherence Times.**

| Parameter | Trion | Exciton |
|---|---|---|
| Valley Coherence, $\tau_v$ ($\gamma_v$) | 230 fs (2.9 meV) | 140 fs (4.8 meV) |
| Valley Polarization | 3 ps (0.22 meV) | 350 fs (1.9 meV) |
| Recombination, $T_1$ ($\Gamma$) | 4.7 ps (0.14 meV) | 210 fs (3.2 meV) |
| Intra-valley Optical Coherence, $T_2$ ($\gamma$) | 510 fs (1.3 meV) | 470 fs (1.4 meV) |
| Intra-valley Pure Dephasing, $T_2^*$ ($\gamma^*$) | 540 fs (1.23 meV) | -- |

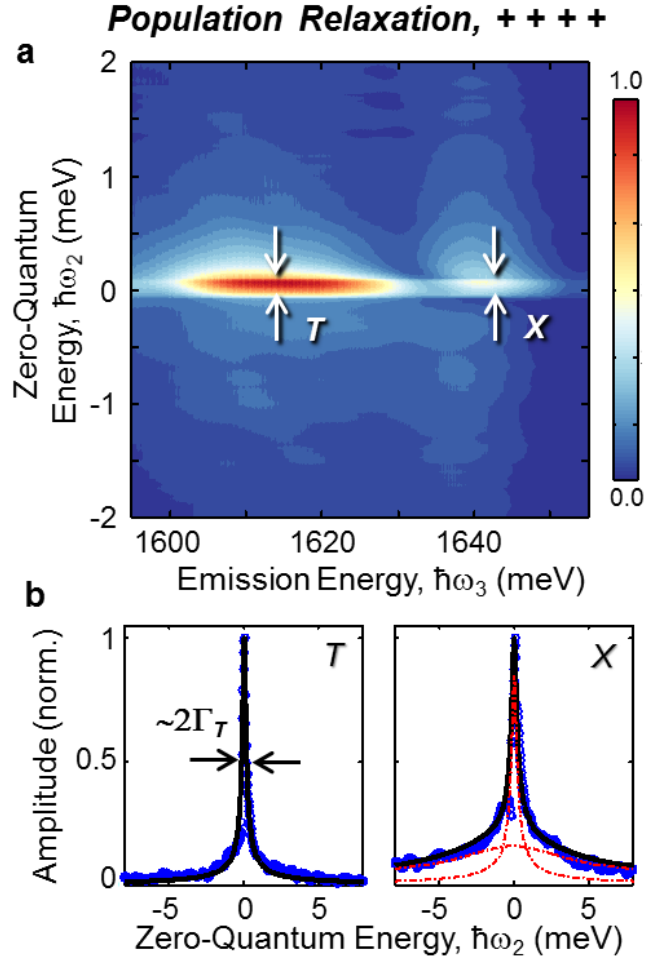

**Figure S1: Population relaxation dynamics of excitons and trions in monolayer MoSe$_2$.** **(a)** Optical 2DCS zero-quantum spectrum acquired using a co-circular polarization scheme. **(b)** The half-width at half-maximum of Lorentzian fit (solid lines) to the exciton (*X*) and trion (*T*) resonance lineshapes along the zero-quantum energy axis ℏ$\omega_2$ provides a measure of the population relaxation rate $\Gamma$ (population lifetime $T_1 = \hbar/\Gamma$). The exciton exhibits a double Lorentzian (bi-exponential) response that is dominated by the wide (fast) component.

## Supplementary Information References